# Structure of visible and dark matter components of spiral galaxies at z ≈ 0.9

**Antti Tamm**[*]

*Tartu University*
*Tähe 4, Tartu, 51010, Estonia*
*E-mail:* `atamm@ut.ee`

**Peeter Tenjes**

*Tartu University*
*Tähe 4, Tartu, 51010, Estonia*
*Tartu Observatory,*
*Tõravere, Tartumaa, 61602, Estonia*
*E-mail:* `ptenjes@ut.ee`

We construct self-consistent light and mass distribution models for 4 distant spiral galaxies. The models include a bulge, a disk and an isothermal dark matter. We find the luminosity profiles to have much steeper cut-off than that of a simple exponential disk. We apply k-corrections and derive rest-frame B-band mass-to-light ratios of the visible components and the central densities of the dark halos; we discover no significant evolution with redshift of these parameters.

[*] Speaker

We have combined the best available observations of photometry and kinematics to construct self-consistent models for spiral galaxies at redshift $z \approx 0.9$, in order to see if any evolution of the parameters of the galaxies is detectable.

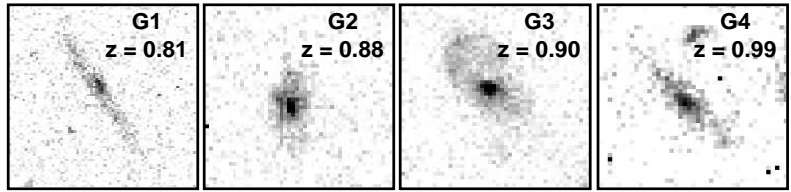

**Fig. 1** Hubble Space Telescope WFPC2 images of galaxies G1-G4 in $I_{815}$ passband

We have used high-resolution Hubble Space Telescope WFPC2-camera $V_{606}$ and $I_{814}$ images to analyze the photometry of 4 distant spiral galaxies (Fig. 1). We have applied image combination, cosmic-ray removal, filtering and point-spread-function deconvolution and finally isophote fitting to the images. We have applied k-correction (according to [1]) to the resulting luminosity profiles to derive rest-frame B-band luminosities (Fig.2).

We have used Keck and VLT rotation curves from papers [2] and [3] to analyze the kinematics of the galaxies. We have refolded the original rotation curves to reduce scatter between the two arms (Fig. 3).

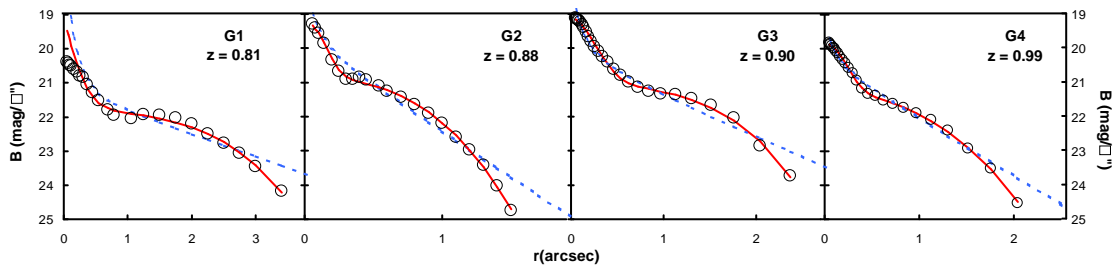

**Fig. 2** Observed luminosities fitted by two different models: blue dotted line – $r^{1/4}$ bulge + exponential disk; red solid line – modified Sersic spatial distribution for both components. To avoid confusion the individual components are not plotted.

We have used two types of models for the visual components: a) exponential spatial distribution $\rho(a) = \rho(0)\exp\{-(a/ka_0)^{1/N}\}$ for both the bulge and the disk; b) the usual surface density $r^{1/4}$- bulge and exponential disk $I(a) = I(0)\exp(r/r_d)$. In both cases isothermal dark matter halos are used. In the first case the density distributions for the bulge and the disk were projected along the line of sight and divided by their mass-to-light ratios. Their sum gives the surface brightness distribution of the model. The masses were determined from the rotation law. If available, also other kinds of information can be easily added to the model, e.g. dispersion analysis, while maintaining its self-consistency. In the second case kinematics can be derived using Bessel functions. However, Fig. 2 shows that the more flexible distribution (a) enables a much better fit with the observations.

The luminosity distribution of all the observed galaxies clearly deviates from the simple exponential disk assumption. The sharp cut-off at outer radii (i.e. the small value of parameter *N*) could be an indication of disk truncation detected for more near-by edge-on spirals (e.g. [4]), but in our case it occurs at somewhat smaller radii.

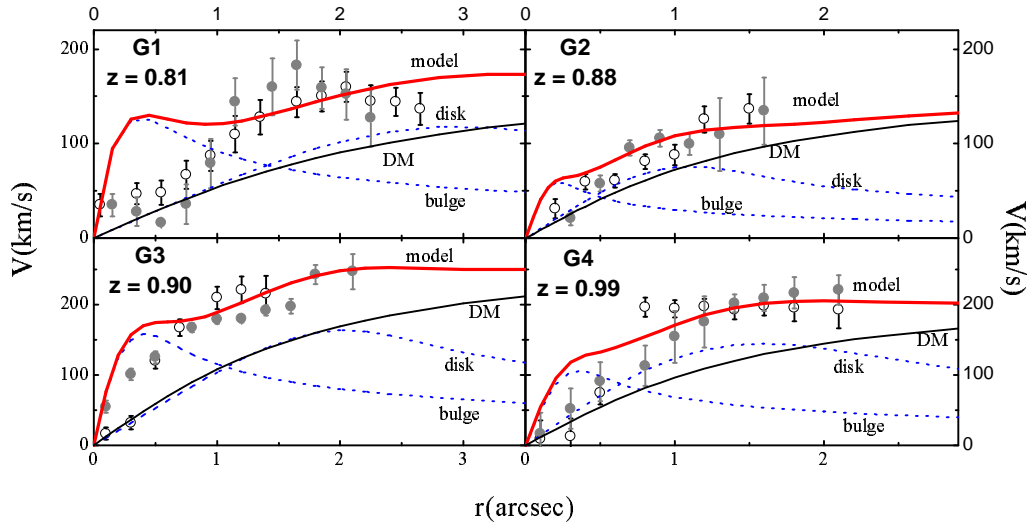

**Fig. 3** Rotation curve observations, models and model components.

The calculated mean mass-to-light ratio at redshift $z = 0.9$ is $M/L_B = 2.5$ $M_\odot/L_\odot$. For local Sb galaxies the value $M/L_B = 2.7$ $M_\odot/L_\odot$ has been determined [5], so we find no significant evolution here.

The calculated mean central density of the dark halos $\rho(0) = 0.02$ $M_\odot/pc^3$ at $z = 0.9$. The local value has been found to range 0.015-0.05 $M_\odot/pc^3$ [6], indicating again no detectable evolution.

Details of the present work can be found in [7] and [8].